\documentclass[a4paper]{jpconf}
\usepackage{graphicx}
\usepackage{amsmath}
\usepackage{amsfonts}
\usepackage{amssymb}
\usepackage{amsbsy}

\newcommand{\del}{\partial}
\newcommand{\intx}{\int\!\mathrm{d}^3x}

\newcommand{\intxx}{\int\!\mathrm{d}^4x\,}
\newcommand{\ssst}{\scriptscriptstyle}

\begin{document}
\title{Treating the Einstein-Hilbert action as a higher derivative Lagrangian: revealing the missing information about conformal non-invariance}

\author{Branislav Nikoli\'c}

\address{Institute for Theoretical Physics, Z\"ulpicher Stra{\ss}e 77, Universit\"at zu K\"oln, 50937 K\"oln, Germany}

\ead{nikolic@thp.uni-koeln.de}

\begin{abstract}
The Hamiltonian formulation of conformally invariant Weyl-squared higher derivative theory teaches us that conformal symmetry is expressed through particular first class constraints related to the absence of the three-metric determinant and the trace of the extrinsic curvature from the theory. Any term depending on them which is added to this theory breaks conformal invariance and turns these constraints into second class ones. Such second class constraints are missing in the standard canonical formulation of the conformally non-invariant Einstein-Hilbert theory. It is demonstrated that such constraints do appear if the theory is treated as a higher derivative one --- if the extrinsic curvature is promoted to an independent variable, the apparently missing information about conformal behavior is revealed.
\end{abstract}

\section{Which variables encode conformal transformation in a theory?}

Conformal (Weyl rescaling) symmetry refers to invariance of a Lagrangian under the following transformation of the metric and the involved fields,
\begin{equation}
\label{eqn:conftrans}
g_{\mu\nu}(x)\quad\rightarrow\quad \tilde{g}_{\mu\nu}(x) = \Omega^2 (x)g_{\mu\nu}(x)\,, \qquad \phi (x)\quad\rightarrow\quad \tilde{\phi}=\Omega^{n} (x)\phi(x)\,,
\end{equation}
where $n$ is usually called the ``conformal weight'' of any kind of field $\phi$. Conformal transformation and related invariance are important concepts. They are studied in many fields of physics, along with symmetry under a conformal transformations of \textit{coordinates} (which we will not talk about here). In particular, while studying theories of gravity, it plays a crucial role e.g. in the transformation between Jordan and Einstein frames in scalar-tensor theories of gravity \cite{CF11}, in renormalization procedures while studying effective theories of gravity \cite{BOS, CS}, as well as in understanding the interplay between the scale and conformal invariance \cite{Nak}. The list is far from complete, but it gives one a taste of the application and importance of the conformal transformation and invariance.

The usual way of studying conformal features of a theory is to directly transform all the involved objects under the conformal transformation \eqref{eqn:conftrans}. The necessary calculations may become quite cumbersome and can either lead to an appearance of non-vanishing terms depending on $\Omega$ and its derivatives, or can lead back to the very same form of the theory in question, in which case all dependence on $\Omega$ cancels (in general, this is true up to a total divergence which might appear in the Lagrangian after transformation). In the latter case, one concludes that the theory is invariant under conformal transformation (up to a total divergence).

Let us demonstrate this on the example of a non-minimally coupled scalar field (the dimension will be assumed to be four, unless otherwise specified):
\begin{equation}
\label{eqn:Lagnm}
\mathcal{L}^{\varphi}=-\frac{1}{2}\sqrt{-g}\left(g^{\mu\nu}\del_{\mu}\varphi\del_{\nu}\varphi+\xi R\varphi^2\right)\,,
\end{equation}
where $\xi$ is a dimensionless non-minimal coupling constant. For the scalar field, we will assume that $n=-1$, since in that case the kinetic term is unchanged after the conformal transformation. Studying conformal properties of this theory requires looking at the transformation of $R$, which is induced by the transformation of 10 variables of the metric; this transformation can be found in the literature,
\begin{equation}
\label{eqn:R-conf}
R\quad\rightarrow\quad \tilde{R} = \frac{1}{\Omega^{2}}\left(R-6\,\frac{\Omega^{-1}}{\sqrt{-g}}\del_{\mu}\left(\sqrt{-g}g^{\mu\nu}\del_{\nu}\Omega\right)\right)\,.
\end{equation}
The resulting transformed Lagrangian reads:
\begin{align}
\label{eqn:Lagnm-trans}
\mathcal{L}^{\varphi}\quad\rightarrow\quad \tilde{\mathcal{L}}^{\varphi}&=-\frac{1}{2}\Biggl[\sqrt{-g}\,g^{\mu\nu}\del_{\mu}\varphi\del_{\nu}\varphi +(1-6\xi)\del_{\mu}\left(\sqrt{-g}\, g^{\mu\nu}\del_{\mu}\Omega\right)\varphi^2+\xi \sqrt{-g}R\varphi^2\Biggr]\nonumber\\[6pt]
&\quad+\frac{1}{2}\del_{\nu}\left(\Omega^{-1}\,\varphi^2\sqrt{-g}\, g^{\mu\nu}\del_{\mu}\Omega\right)\,.
\end{align}
The second term spoils conformal invariance. The piece proportional to $-6\xi$ comes from the non-minimal coupling term $\xi R\varphi^2$, while the piece proportional to $1$ comes from the kinetic term, using the Leibniz rule (giving rise to the last term). We see that the suggestive value $\xi = 1/6$ exactly cancels the $\Omega-$dependence from the original kinetic term. This is called conformal coupling. Hence one says that the Lagrangian of the conformally coupled massless scalar field is conformally invariant up to a total divergence if the scalar field transforms as $\tilde{\varphi}=\Omega^{-1}\varphi$.

It will be demonstrated now that an appropriate decomposition of the variables suggests that \textit{only one} of the resulting configuration variables is responsible for the conformal transformation of the Lagrangian --- the determinant of the metric (raised to an appropriate power). This decomposition is deduced from the following reasoning. The metric has 10 independent elements, i.e., 10 degrees of freedom to start with. Its determinant is an object which carries only one degree of freedom --- it is a scalar density. As a result of \eqref{eqn:conftrans}, the determinant transforms as
\begin{equation}
\label{eqn:det-transf}
\sqrt{-g}\quad\rightarrow\quad \sqrt{-\tilde{g}}=\Omega^4\sqrt{-g}
\end{equation}
in four dimensions. We can then say that a single power of $\Omega$ is produced from each $(\sqrt{-g})^{1/d}$. In this way, the conformal transformation of the metric can be thought of as if it results from a single degree of freedom only --- the determinant. It is thus suggestive to decompose the metric into its determinant part and the remaining 9 degrees of freedom. This is achieved in the following way:
\begin{equation}
\label{eqn:umod-dec}
A:=\left(\sqrt{-g}\right)^{1/d}\,,\qquad g_{\mu\nu}=A^2\bar{g}_{\mu\nu}\,,\qquad g^{\mu\nu}=A^{-2}\bar{g}^{\mu\nu}\,,\qquad  {\rm det}\bar{g}_{\mu\nu} = 1
\end{equation}
in $d$ dimensions, where $\bar{g}_{\mu\nu}$ is usually called ``unimodular metric'', due to its unit determinant (which is why it has $d-1$ degrees of freedom), or ``conformal part of the metric'', since it is invariant under conformal transformation. This is an \textit{irreducible} decomposition under conformal (Weyl rescaling) transformation into its \textit{scale density} (non-conformal) $A$ and unimodular (conformal) $\bar{g}_{\mu\nu}$ part. Hence, it can now be understood that the conformal transformation of the metric results from the conformal transformation of its scale density $(A)$ degree of freedom only. But nothing prevents one from extending this logic to \textit{all} variables appearing in a Lagrangian. Namely, conformal transformation of the scalar field suggests the following decomposition
\begin{equation}
\label{eqm:scal-dec}
\varphi = A^{-1}\chi\,,
\end{equation}
where $\chi$ is a \textit{conformally invariant scalar density of weight} $1/d$. Hence, the \textit{geometric} degree of freedom (responsible for the conformal transformation) is separated from what one may call the ``pure'' matter degree of freedom, the scalar density $\chi$.

With this decomposition, one may reinterpret the conformal behavior of the Lagrangian \eqref{eqn:Lagnm-trans}. Namely, decomposition of the Ricci tensor induced by the unimodular decomposition of the four-metric results in (see Appendix in \cite{KN17c})
\begin{align}
\label{eqn:RicciScalDec1}
 R&=\frac{1}{A^2}\Biggl(\bar{R}-\frac{2(d-1)}{A^2}\biggl[A\,\del_{\mu}\left(\bar{g}^{\mu\nu}\del_{\nu}A\right)+\frac{(d-4)}{2}\bar{g}^{\mu\nu}\del_{\mu}A\,\del_{\nu}A\biggr]\Biggr)\nonumber\\[6pt]
&\stackrel{d=4}{=}\frac{1}{A^2}\Biggr(\bar{R}-\frac{6}{A}\,\del_{\mu}\left(\bar{g}^{\mu\nu}\del_{\nu}A\right)\Biggr)\,,
\end{align}
where $\bar{R}$ is the part of the Ricci scalar determined only by the unimodular part of the metric, and is therefore conformally invariant. The conformal transformation of the Ricci scalar is now obtained simply by transforming $A\rightarrow \Omega A$ only. It is easy to see that
\begin{equation}
\frac{1}{A^3}\,\del_{\mu}\left(\bar{g}^{\mu\nu}\del_{\nu}A\right)\quad\rightarrow\quad \frac{1}{\Omega^2 A^3}\,\del_{\mu}\left(\bar{g}^{\mu\nu}\del_{\nu}A\right)+\frac{\Omega^{-3}}{\sqrt{-g}}\del_{\mu}\left(\sqrt{-g}g^{\mu\nu}\del_{\nu}\Omega\right)
\end{equation}
indeed gives rise to the conformal transformation \eqref{eqn:R-conf} already known from the literature (the last term is expressed in terms of original variables, for reference). Furthermore, applying the decomposition to \eqref{eqn:Lagnm}, we obtain the Lagrangian
\begin{align}
\label{eqn:Lagnm-decA}
\mathcal{L}^{\varphi}=\mathcal{L}^{\chi}&=-\frac{1}{2}\Biggl[\bar{g}^{\mu\nu}\del_{\mu}\chi\del_{\nu}\chi+(1-6\xi)\del_{\mu}\left(\, \bar{g}^{\mu\nu}\del_{\mu}A\right)\chi^2+\xi \,\bar{R}\chi^2\Biggr]+\frac{1}{2}\del_{\nu}\left(A^{-1}\,\chi^2\, \bar{g}^{\mu\nu}\del_{\mu}A\right)\,,
\end{align}
reformulated for the non-minimally coupled scalar \textit{density} field $\chi$. An easy calculation shows that $A\rightarrow \Omega A$ reproduces the known conformal behavior \eqref{eqn:Lagnm-trans}. One can now understand the conformal features of this potentially conformally invariant Lagrangian in terms of the dependence on the scale density $A$. It is a simple matter to deduce that $A$ is completely eliminated (up to a total divergence) only if $\xi = 1/6$, which is the conformal coupling, resulting in the conformally invariant Lagrangian.

The lesson to take away from this introductory section is the following. Suppose a given theory is formulated terms of the scale density and conformally invariant variables; it follows that if the theory does not depend on scale density (up to a total divergence), it is conformally invariant.

\section{Which constraints encode conformal features of a theory?}

How can we learn of conformal behavior of theories in the Hamiltonian formulation? In studying the Hamiltonian formulation of metric theories of gravity, one usually uses the 3+1 decomposition of spacetime \cite{Pad}, such that the central role is taken over by the three-metric, instead of the four metric. Namely, the four-metric is decomposed as
\begin{equation}
\label{eqn:g31mat}
g_{\mu\nu}=\left(
\begin{matrix}
-N^{2}+N_{i}N^{i}& N_{i}\\[6pt]
N_{i} & h_{ij} &
\end{matrix}
\right)\,,\quad \sqrt{-g}=N\sqrt{h}\,,
\end{equation} 
where $N$ and $N^{i}$ are the lapse and the shift functions, and $h_{ij}$ is the three-metric, the metric intrinsic to the 3-dimensional hypersurface, with its determinant denoted by $h$. Of importance is another tensor, called the extrinsic curvature, defined by
\begin{equation}
\label{eqn-Ks}
K_{ij}=\frac{1}{2N}\left(\dot{h}_{ij}-2D_{(i}N_{j)}\right)\,,\quad
K=h^{ij}K_{ij}=\frac{1}{N}\left(\frac{\dot{\sqrt{h}}}{\sqrt{h}}-D_{i}N^{j}\right)\,,
\end{equation}
where $K$ is its trace (note that it represents time evolution of the three-volume $\sqrt{h}$), and $D_{i}$ is the spatial covariant derivative. Equipped with these tensors, one can canonically formulate any metric theory of gravity via what is known as the ADM formalism (see e.g. \cite{Pad}). However, for purposes discussed here, guided by the lesson from the previous section, let us go a step further and decompose the $3+1$ variables in such a way that their conformal transformation is completely determined only by the transformation of the \textit{three-dimensional scale density} (the determinant of the three-metric $\sqrt{h}$), in analogy to the full four-dimensional case, discussed in the previous section. These variables are introduced in \cite{KN17c} (see also \cite{KN17D}) and are defined by,
\begin{align}
\label{eqn:Vars-m}
& a :=(\sqrt{h})^{1/3}\,,\quad\bar{h}_{ij}:=a^{-2}h_{ij}\,,\quad\chi := a \varphi\\[6pt]
\label{eqn:VarsNbar}
&\bar{N}^{i}:=N^{i}\,,\quad\bar{N}_{i}:=a^{-2} N_{i}\,,\quad\bar{N}:=a^{-1}N\,,\\[6pt]
\label{eqn:VarsKbar}
& \bar{K}_{ij}^{\ssst\rm T}:=a^{-1}{K}_{ij}^{\ssst\rm T}=\frac{1}{2\bar{N}}\left(\dot{\bar{h}}_{ij}-2\left[\bar{D}_{(i}\bar{N}_{j)}\right]^{\ssst\rm T}\right)\,,\quad\,\,\,
\bar{K}:=\frac{aK}{3}=\frac{1}{\bar{N}}\left(\frac{\dot{a}}{a}-\frac{1}{3}D_{i}N^{i}\right)\,,
\end{align}
where $\bar{h}_{ij}$ is now our unimodular part of the metric, and $a$ is the three-dimensional scale density, in analogy with \eqref{eqn:umod-dec}. Note that $N^{i}$ is the only unchanged variable. Superscript ``$\ssst\rm T$'' denotes the traceless object, $(i...j)$ represent symmetrization of indices, while $\bar{D}_{i}$ acts only with the conformal part of the three-dimensional covariant derivative. They may be called ``unimodular-conformal variables''. We have added the matter scalar density $\chi$ for completeness. It can be easily checked that only two of the above variables transform under a conformal transformation: the scale density $a$ and the ``trace density'' $\bar{K}$. The trace density $\bar{K}$ is the only variable determined by the scale density $a$ (through $K$, see \eqref{eqn-Ks}), and contains its temporal evolution, i.e. its velocity. Therefore, we expect that the conformal properties of any theory may be expressed in terms of the scale density $a$ and the trace density $\bar{K}$, and so we keep an eye on them throughout the derivations.

The Einstein-Hilbert (EH) Lagrangian is usually given in the ADM form, but we further reformulate it in terms of variables \eqref{eqn:Vars-m}-\eqref{eqn:VarsKbar} to obtain,
\begin{equation}
\label{eqn:E-Lag}
\mathcal{L}^{\ssst\rm E}=\frac{1}{2\kappa} N\sqrt{h}\left(\,^{\ssst (3)}\! R+K_{ij}K^{ij}-K^2\right)=\frac{1}{2\kappa}\bar{N}a^2\left(a^2\,^{\ssst (3)}\! R+\bar{K}_{ij}^{\ssst\rm T 2}-6\bar{K}^2\right)\,,
\end{equation}
where $\kappa \equiv 8\pi G$, $\bar{K}_{ij}^{\ssst\rm T 2}\equiv \bar{K}_{ij}^{\ssst\rm T}\bar{h}^{ia}\bar{h}^{jb}\bar{K}_{ab}^{\ssst\rm T}$, and $\,^{\ssst (3)}\! R$ is left undecomposed for simplicity, but recall that it contains a dependence on $a$. It is now obvious that $a$ and $\bar{K}$ are responsible for conformal transformations of the EH Lagrangian. It is well known \cite{Pad} that the canonical EH theory is a \textit{constrained} theory, meaning that it contains constraint equations. These constraints are the consequence of $\dot{N}$ and $\dot{N}^{i}$ not appearing in the Lagrangian, which gives rise to their vanishing momenta as primary constraints. Time preservation of these constraints gives rise to the secondary constraints called Hamiltonian $\mathcal{H}_{\bot}$ and momentum constraint $\mathcal{H}_{i}$, respectively. In unimodular-conformal variables, it is the absence of $\dot{\bar{N}}$ and $\dot{\bar{N}}^{i}$ which gives rise to the primary-secondary pair of constraints, having the following form,
\begin{align}
\label{eqn:EH-pH1}
p_{\ssst \bar{N}}&\approx 0 \quad\Rightarrow\quad \mathcal{H}^{\ssst\rm E}_{\bot}= -\frac{\kappa}{12}\, p_{a}^{2}+\frac{2\kappa}{a^2}\bar{h}_{ik}\bar{h}_{jl}\bar{p}^{ij}\bar{p}^{kl}\nonumber\\[6pt]
&\qquad\qquad\qquad\quad\,\,-\frac{a^2}{2\kappa}\left(\,^{\ssst (3)}\! \bar{R}-\frac{4}{a^2}\biggl[a\,\del_{i}\left(\bar{h}^{ij}\del_{j}a\right)-\frac{1}{2}\bar{h}^{ij}\del_{i}a\,\del_{j}a\biggr]\right)\approx 0\,,\\[6pt]
\label{eqn:EH-pH2}
p_{i}&\approx 0 \quad\Rightarrow\quad \mathcal{H}^{\ssst\rm E}_{i}=-2\del_{k}\left(
\bar{h}_{ij}\bar{p}^{jk}\right)+\del_{i}\bar{h}_{jk}\bar{p}^{jk}-\frac{1}{3}D_{i}\left(a\,p_{a}\right)\approx 0\,,
\end{align}
where the momenta conjugate to the decomposed parts of the metric are
\begin{equation}
\label{eqn:E-ps}
p_{a}=\frac{\del \mathcal{L}^{\ssst\rm E}}{\del \dot{a}}=-\frac{6a}{\kappa}\bar{K}\,,\quad \bar{p}^{ij}=\frac{\del \mathcal{L}^{\ssst\rm E}}{\del \dot{\bar{h}}_{ij}}=\frac{a^2}{2\kappa}\bar{K}_{ab}^{\ssst\rm T}\bar{h}^{ia}\bar{h}^{jb}\,,
\end{equation}
and can be shown to be related to the trace and traceless parts of the ADM momentum $p^{ij}=\sqrt{h}\left(K^{ij}-h^{ij}K\right)/(2\kappa)$. The total Hamiltonian is given by
\begin{equation}
\label{eqn:E-totHam1}
H^{\ssst\rm E}=\intx\left\lbrace \bar{N}\mathcal{H}^{\ssst\rm E}_{\bot}+N^{i}\mathcal{H}^{\ssst\rm E}_{i}+\lambda_{\ssst\bar{N}}p_{\ssst\bar{N}}+\lambda^{i}p_{i}\right\rbrace\,.
\end{equation}
Note that we have decomposed the Ricci scalar above as well \cite{KN17c}, in order to identify all terms where the scale density $a$ resides. The sign ``$\approx $'' is Dirac's ``weak equality'', used in the constraint analysis to enable evaluation of the Poisson brackets. The only feature of canonical GR of importance for the present discussion is the fact that the constraints \eqref{eqn:EH-pH1} and \eqref{eqn:EH-pH2} are \textit{first class} constraints. This means that they are related to a certain symmetry of the theory, in this case it is the covariance of the Lagrangian. Whenever one encounters first class constraints, it is a signal that a symmetry is present in the theory, and vice versa. It is now important to keep in mind that these are \textit{the only} constraints in the canonical EH theory.

Apart from GR, some further examples of Hamiltonian formulation of such theories can be found in e.g. \cite{KN17c, Kaku1982, Blw, DerrHD, Kluson2014}, which contain curvature-squared terms $R^2, R_{\mu\nu}^2, R_{\mu\nu\rho\sigma}^2$. These theories are considered either as an extension of GR or as an alternative to it, from both the classical and quantum point of view, and are usually referred to as ``higher derivative theories'', because they are non-linear in second order time derivatives of the metric. Hence, unlike pure GR, the order cannot be reduced by simple partial integration and using a boundary term. One then usually promotes the extrinsic curvature to an \textit{independent} variable, by adding a constraint to the Lagrangian,
\begin{align}
\label{eqn:Lagc}
\mathcal{L} (N, N^i , h_{ij},\dot{h}_{ij},\ddot{h}_{ij})\quad\rightarrow\quad\mathcal{L}'(N, N^i ,h_{ij},K_{ij},\dot{K}_{ij})-N\sqrt{h}\lambda^{ij}\left[2K_{ij}-\frac{1}{N}\left(\dot{h}_{ij}-2D_{(i}N_{j)}\right)\right]\,,
\end{align}
where $\mathcal{L}'$ is the same Lagrangian $\mathcal{L}$, but expressed in terms of new variables $K_{ij}$. This ``hides'' the second derivatives $\ddot{h}_{ij}$ and one can then proceed with the Hamiltonian formulation. The added constraint with Lagrange multiplier $\lambda^{ij}$ follows from \eqref{eqn-Ks} and takes care of the additional number of degrees of freedom. It is important to realize that the momentum conjugate to the metric components is $p^{ij}=\sqrt{h}\lambda^{ij}$, i.e., is undetermined ($\lambda^{ij}$ can be eliminated from the configuration space from the start, see \cite{Kluson2014}). Using unimodular-conformal variables, it is possible to further reformulate such theories, leading to a great insight into their conformal features expressed in terms of $a$ and $\bar{K}$. As an example, we will take the Weyl-squared theory, whose Hamiltonian formulation is investigated in detail in \cite{KN17c}, along with some of its quantum aspects in \cite{KN17D},
\begin{equation}
\label{eqn:W-lag}
\mathcal{L}^{\ssst \rm W}:=-\frac{\alpha_{\ssst\rm W}}{4}\intxx\sqrt{-g}\,C_{\mu\nu\lambda\rho}C^{\mu\nu\lambda\rho}\,.
\end{equation}
Here, $C^{\mu}_{\,\,\,\,\nu\lambda\rho}$ is the Weyl tensor, and $\alpha_{\ssst\rm W}$ is a positive dimensionless coupling constant. Due to covariance, the Hamiltonian and momentum constraints are present \cite{DerrHD}, as in GR, but will not be discussed here (see \cite{KN17c} for more details). However, \eqref{eqn:W-lag} enjoys conformal symmetry due to the conformal invariance of the Weyl tensor, and this richer structure is the reason why we choose this particular example. Consequentially, this theory generates \textit{additional} first class constraints related to the generator of gauge conformal transformations, as first determined by \cite{ILP}. They arise because $a$ and $\bar{K}$ do not appear in the Weyl-squared theory --- an important fact which seems not to have been completely realized in the literature until the theory was treated in unimodular-conformal vriables in \cite{KN17c}. These constraints read
\begin{align}
\label{eqn:Qw}
\bar{P}=\frac{\del \mathcal{L}^{\ssst\rm W}}{\del \dot{\bar{K}}}\approx 0\quad\Rightarrow\quad\dot{\bar{P}}=\mathcal{Q}^{\ssst\rm W}=ap_{a}\approx 0\,,
\end{align}
whose Poisson bracket trivially vanishes. The constraint $\bar{P}\approx 0$ appears because the Lagrangian does not depend on $\dot{\bar{K}}$, and as a consequence, $\mathcal{Q}^{\ssst\rm W}$ effectively states the same for $\dot{a}$, thanks to the fact that the Lagrangian is independent of $\bar{K}$. It is important to keep in mind that the appearance of these additional constraints is intimately related to the treatment of the extrinsic curvature as an \textit{independent variable}, in contrast to the Lagrangian formulation of GR, where $K_{ij}$ is not a configuration variable.

We are now interested the cahnge of constraints \eqref{eqn:Qw} if some non-conformal term is added to the Weyl-squared action. Since this is studied in detail in \cite{KN17c}, here we borrow only the results. Namely, if the added term is the Lagrangian of the non-minimally coupled scalar field \eqref{eqn:Lagnm}, the Hamiltonian analysis shows that constraints \eqref{eqn:Qw} are turned into second class constraints and read
\begin{equation}
\label{eqn:Qwnm}
\bar{P}\approx 0\quad\Rightarrow\quad
\mathcal{Q}^{{\ssst\rm W}\chi}=ap_{a}-6\xi\left(1-6\xi\right)a^{2(1-6\xi)}\bar{K}\chi^2\approx 0\,,
\end{equation}
with their non-weakly-vanishing Poisson bracket $\left\lbrace \bar{P},\,\mathcal{Q}^{{\ssst\rm W}\chi}\right\rbrace=6\xi\left(1-6\xi\right)a^{2(1-6\xi)}\chi^2$ (we omit the Dirac delta function). It is quite clear that the first class nature of these constraints is recovered only if $\xi=1/6$, i.e., only if the scalar field is conformally coupled (if $\xi = 0$ the matters are more subtle, since $\chi = a^{6\xi}\varphi$ was used in \cite{KN17c}). Looking more closely, we see that $\xi$ controls the appearance of the scale density $a$ and trace density $\bar{K}$, and we can conclude that the appearance of $a$ and $\bar{K}$ in the Lagrangian is responsible for the breaking of conformal symmetry. Thus, the information about conformal features is \textit{not lost} upon breaking of symmetry --- it is still contained in the \textit{second class} constraints.

If that is the case, then where is the information about the broken conformal symmetry in the canonical EH theory? Recall that there one has only first class constraints \eqref{eqn:EH-pH1}-\eqref{eqn:EH-pH2}. Where are the second class constraints expressing its conformal non-invariance? To answer this question, one has to realize that any additional constraints would further reduce the number of degrees of freedom in GR, which we know is two. Thus, additional constraints cannot appear unless additional variables are introduced to the theory. On the other hand, we have the Weyl-squared theory whose configuration space necessarily has to be expanded by additional variables, i.e. the extrinsic curvature, and we have seen that these variables ($\bar{K}$ in particular) gave rise to additional constraints related to the conformal (non-)invariance. Let us then combine the two observations and propose to treat the EH theory as if it were a higher derivative theory, and see if we can learn something about its conformal behavior.

\section{General Relativity treated as a higher derivative theory: manifest absence of conformal non-invariance}

The EH Lagrangian does not contain second order time derivatives of the three-metric (after the partial integration), and it is not necessary to introduce the extrinsic curvature as an independent variable. But it is not harmful, either. Let us take for the Lagrangian $\mathcal{L}$ in \eqref{eqn:Lagc} to be the EH Lagrangian \eqref{eqn:E-Lag} and call it $\mathcal{L}^{\ssst\rm EH}$. After some calculation similar to that in \cite{KN17c}, the total Hamiltonian can be put in the following form,
\begin{equation}
\label{eqn:E-totHam}
H^{\ssst\rm EH}=\intx\left\lbrace \bar{N}\mathcal{H}^{\ssst\rm EH}_{\bot}+N^{i}\mathcal{H}^{\ssst\rm EH}_{i}+(2\bar{N}\bar{K}_{ij}^{\ssst\rm T})\mathcal{T}^{ij}_{\ssst\rm EH}+\left(\bar{N}\bar{K}\right)\mathcal{Q}^{\ssst\rm EH}+\lambda_{\ssst\bar{N}}p_{\ssst\bar{N}}+\lambda^{i}p_{i}+\Lambda_{ij}\bar{P}^{ij}+\Lambda_{\ssst\bar{K}}\bar{P}\right\rbrace\,,
\end{equation}
where $\bar{N},N^{i},(2\bar{N}\bar{K}_{ij}^{\ssst\rm T}),\left(\bar{N}\bar{K}\right),\lambda_{\ssst\bar{N}},\lambda^{i},\Lambda_{ij},\Lambda_{\ssst\bar{K}}$ are Lagrange multipliers, and the following are constraints:
\begin{align}
\label{eqn:E-HamCf}
p_{\ssst \bar{N}}&\approx 0 \quad\Rightarrow\quad\mathcal{H}^{\ssst\rm EH}_{\bot}=-\frac{a^2}{2\kappa}\biggl(\,^{\ssst (3)}\bar{R}-\bar{K}_{ij}^{{\ssst\rm T}2}+6\bar{K}^2\biggr)+\frac{2}{\kappa}\biggl[a\,\del_{i}\left(\bar{h}^{ij}\del_{j}a\right)-\frac{1}{2}\bar{h}^{ij}\del_{i}a\,\del_{j}a\biggr]\approx 0\,,\\[6pt]
\label{eqn:E-MomCf}
p_{i}&\approx 0 \quad\Rightarrow\quad\mathcal{H}^{\ssst\rm EH}_{i}=-2\del_{k}\left(
\bar{h}_{ij}\bar{p}^{jk}\right)+\del_{i}\bar{h}_{jk}\bar{p}^{jk}-\frac{1}{3}D_{i}\left(a\,p_{a}\right)\approx 0\,,\\[6pt]
\label{eqn:E-Qkf}
\bar{P}&\approx 0 \quad\Rightarrow\quad\mathcal{Q}^{\ssst\rm EH}\equiv ap_{a}+\frac{6a^2}{\kappa}\bar{K}\approx 0\,,\\[6pt]
\label{eqn:E-Tijf}
\bar{P}^{ij}&\approx 0 \quad\Rightarrow\quad\mathcal{T}^{ij}_{\ssst\rm EH}\equiv 2\bar{p}^{ij}-\frac{a^2}{\kappa}\bar{K}_{ab}^{\ssst\rm T}\bar{h}^{ai}\bar{h}^{bj}\approx 0\,.
\end{align}
Compared to the constraints \eqref{eqn:EH-pH1} and \eqref{eqn:EH-pH2} from standard canonical GR, there are a few striking differences (apart from the momentum constraint \eqref{eqn:E-MomCf}, which is the same as \eqref{eqn:EH-pH2}). First of all, there is no kinetic term in the Hamiltonian constraint \eqref{eqn:E-HamCf}. The momentum conjugate to the three-metric (and its decomposed parts) is undetermined through the Lagrange multiplier $\lambda^{ij}$ from \eqref{eqn:Lagc}, see discussion which followed. Even though this might seem alarming, it is still the EH Lagrangian we are dealing with and no choice of variables should affect the dynamical structure of the theory. This leads us to investigate the second observation. Namely, there are six additional constraints \eqref{eqn:E-Qkf} and \eqref{eqn:E-Tijf}, which turn out to be pairs of \textit{second class constraints}, because they do not commute,
\begin{equation}
\label{eqn:E-PB2nd}
\left\lbrace \bar{P},\,\mathcal{Q}^{\ssst\rm EH}\right\rbrace=-\frac{6a^2}{\kappa}\,,\quad \left\lbrace \bar{P}^{ij},\,\mathcal{T}^{mn}_{\ssst\rm EH}\right\rbrace=\frac{a^2}{\kappa}\left(\bar{h}^{im}\bar{h}^{jn}-\frac{1}{3}\bar{h}^{ij}\bar{h}^{mn}\right)\,.
\end{equation}
What is the reason for this? Observe that the constraints \eqref{eqn:E-Qkf} arise due to the absence of $\dot{\bar{K}}$ from the EH Lagrangian. A similar situation was encountered in the Weyl-squared theory, where the absence of $\dot{\bar{K}}$ resulted in first class constraints \eqref{eqn:Qw}. However, the difference here is that $\bar{K}$ is \textit{not} absent from the EH Lagrangian as in the Weyl-squared theory \cite{KN17c}, which is why \eqref{eqn:E-Qkf} are second class constraints. Also, additional constraints \eqref{eqn:E-Tijf} are not met in the Weyl-squared theory, because $\bar{K}_{ij}^{\ssst\rm T}$ is a dynamical variable there. These constraints are second class for the similar reason as \eqref{eqn:E-Qkf} --- $\bar{K}_{ij}^{\ssst\rm T}$ cannot be eliminated from the theory. It seems that our new formulation of GR is contaminated with non-dynamical degrees of freedom. However, not all is lost. It turns out that $\dot{\mathcal{Q}}^{\ssst\rm EH}\approx 0$ and $\dot{\mathcal{T}}^{ij}_{\ssst\rm EH}\approx 0 $ determine $\Lambda_{\bar{K}}$ and $\Lambda_{ij}$, respectively, but their explicit form is tedious to calculate and of no direct relevance for our discussion. The important thing is that second class constraints \eqref{eqn:E-Qkf} and \eqref{eqn:E-Tijf} can now be implemented, which does two remarkable things: firstly, it reduces the total Hamiltonian to the following form
\begin{equation}
\label{eqn:E-totHamRed}
H^{\ssst\rm EH}=\intx\left\lbrace \bar{N}\mathcal{H}^{\ssst\rm EH}_{\bot}+N^{i}\mathcal{H}^{\ssst\rm EH}_{i}+\lambda_{\ssst\bar{N}}p_{\ssst\bar{N}}+\lambda^{i}p_{i}\right\rbrace\,;
\end{equation}
which is of the same familiar form as \eqref{eqn:E-totHam1}; and secondly, the seemingly arbitrary $\bar{K}$ and $\bar{K}^{\ssst\rm T}_{ij}$ and seemingly arbitrary $p_{a}$ and $\bar{p}^{ij}$ turn out to be actually related,
\begin{align}
\label{eqn:E-paij}
p_{a}=-\frac{6a^2}{\kappa}\bar{K}\,,\quad \bar{p}^{ij}=\frac{a}{2\kappa}\bar{K}_{ab}^{\ssst\rm T}\bar{h}^{ai}\bar{h}^{bj}\,,
\end{align}
Remarkably, this relationship is nothing than the definitions of the momenta in \eqref{eqn:E-ps} in the standard ADM treatment of GR! Eliminating $\bar{K}^{\ssst\rm T}_{ij}$ and $\bar{K}$ from \eqref{eqn:E-HamCf} via \eqref{eqn:E-paij} in terms of  $p_{a}$ and $\bar{p}^{ij}$ recovers \eqref{eqn:EH-pH1}, which completes the equivalence of the two formulations. Moreover, one may check that the number of physical degrees of freedom is still 2: 16 original configuration variables $\bar{N},N^{i},\bar{h}_{ij},a,\bar{K}^{\ssst\rm T}_{ij}$ and $\bar{K}$, 8 first class constraints \eqref{eqn:E-HamCf} and \eqref{eqn:E-MomCf}, and 6 pairs of second class constraints \eqref{eqn:E-Qkf} and \eqref{eqn:E-Tijf} give $16-8-6 = 2$ physical degrees of freedom, as expected.

\section{Conclusion}

We learn from the Weyl-squared theory that the constraints 
$\bar{P},\mathcal{Q}^{\ssst\rm W}$ are first class only if $a$ and $\bar{K}$ --- the only conformally variant variables --- are absent from the theory. Any dependence of the Lagrangian on $a$ and $\bar{K}$ induces a breaking of conformal symmetry, and turns the constraints into second class ones. In the standard formulation of the EH theory, for which we know is conformally non-invariant, there are no second class constraints which may give information about the obviously broken conformal symmetry. However, if extrinsic curvature is promoted to an independent configuration variable, the momentum conjugate to variable $\bar{K}$ turns out to vanish. This gives rise to \textit{second class} constraints \eqref{eqn:E-Qkf}, similarly to the Weyl-squared theory with non-minimally coupled scalar field \eqref{eqn:Qwnm}. They express the missing information about the broken conformal symmetry of the EH theory. Similarly, the absence of $\dot{K}_{ij}^{\ssst\rm T}$ from the EH Lagrangian gives rise to second class constraints \eqref{eqn:E-Tijf}.

Thus, one may say that broken conformal symmetry turns scale density (the volume) into a \textit{dynamical} variable. The implementation of the second class constraints ensures the equivalence with the standard ADM formulation, by recovering the ADM momentum \eqref{eqn:E-paij}. The question remains, however, what is the meaning of the second class constraints \eqref{eqn:E-Tijf} which give rise to dynamics of the unimodular part of the metric $\bar{h}_{ij}$, but we leave this for another occasion.

\section*{Acknowledgments}

The author would like to thank Claus Kiefer for his encouragement to write this paper and for reading the manuscript. He is also thankful to Anirudh Gundhi, Nick Kwidzinski and David Wichmann for helpful comments.

\end{document}